\documentclass[fleqn,twoside]{article}
\usepackage{espcrc2}
\usepackage{slashed}
\usepackage{slashbox}
\usepackage{multirow}
\usepackage{graphicx}
\usepackage[dvips]{color}
\usepackage{epsfig,psfig}
\usepackage{colortbl}

\usepackage{graphicx}
\usepackage[figuresright]{rotating}

\newcommand{\AmS}{{\protect\the\textfont2
  A\kern-.1667em\lower.5ex\hbox{M}\kern-.125emS}}
\newcommand{\be}{\begin{equation}}
\newcommand{\ee}{\end{equation}}
\newcommand{\ba}{\begin{array}}
\newcommand{\ea}{\end{array}}
\newcommand{\bea}{\begin{eqnarray}}
\newcommand{\eea}{\end{eqnarray}}
\newcommand{\neff}{N_{\rm eff}}
\newcommand{\pp}{~~~.}
\newcommand{\vv}{~~~,}
\def\yp{{\rm Y}_{p}}
\def\He4{{}^4{\rm He}}
\def\he3{{}^3{\rm He}}
\def\Li6{{}^6{\rm Li}}
\def\li7{{}^7{\rm Li}}
\def\gsim{\;\raise0.3ex\hbox{$>$\kern-0.75em\raise-1.1ex\hbox{$\sim$}}\;}
\def\lsim{\;\raise0.3ex\hbox{$<$\kern-0.75em\raise-1.1ex\hbox{$\sim$}}\;}

\newcommand{\PR}{{\it Phys. Rev.\,}}

\title{Primordial Nucleosynthesis: an updated comparison of observational light nuclei abundances with
theoretical predictions}

\author{G. Miele\address[UNINA]{Universit\`a di Napoli ``Federico II'', Dipartimento di
Scienze Fisiche and INFN - Sezione di Napoli, Complesso
Universitario di Monte S.Angelo, Via Cithia, 80126, Napoli, Italy}
\address[IFIC]{Instituto de F\'{\i}sica Corpuscular (CSIC-Universitat de
Val\`encia),
        Ed.\ Institutos de Investigaci\'on, Apartado de Correos 22085, E-46071 Val\`encia,
        Spain.},O. Pisanti\addressmark[UNINA]}

\begin{document}

\begin{abstract}
An up to date review of Standard Big Bang Nucleosynthesis
predictions vs the astrophysical estimates of light nuclei
abundances is here presented. In particular the analysis reports
the expected ranges for baryon fraction and effective number of
neutrinos as obtained by BBN only. \vspace{1pc}
\end{abstract}

\maketitle

\section{Introduction}

The nucleosynthesis taking place in the primordial plasma plays a
twofold role: it is certainly one of the observational pillars of
the hot Big Bang model, being indeed known simply as ``Big Bang
Nucleosynthesis'' (BBN); at the same time, it provides one of the
earliest direct cosmological probe nowadays available,
constraining the properties of the universe when it was a few
seconds old, or equivalently at the MeV temperature scale.

The basic framework of the BBN emerged in the decade between  the
seminal Alpher-Bethe-Gamow (known as $\alpha\beta\gamma$) paper in
1948 \cite{Alp48} and the essential settlement of the paradigm of
the stellar nucleosynthesis of elements heavier than $^7$Li with
the B$^2$FH paper \cite{Bur57}. This pioneering period---an
account of which can be found in \cite{Kra96}---established the
basic picture that sees the four light-elements $^2$H, $^3$He,
$^4$He and $^7$Li as products of the early fireball, and virtually
all the rest produced in stars or as a consequence of stellar
explosions.

In the following decades, the emphasis on the role played by BBN
has evolved significantly.  In the simplest scenario, the only
free parameters in primordial nucleosynthesis are the baryon to
photon ratio $\eta$  (equivalently, the baryon density of the
universe) and the neutrino chemical potentials,
$\mu_{\nu_\alpha}$. However, only neutrino chemical potentials
larger than $\eta$ by many orders of magnitude have appreciable
effects. This is why the simple case where all
$\mu_{\nu_\alpha}$'s  are assumed to be negligibly small (e.g., of
the same order of $\eta$) is typically denoted as Standard BBN
(SBBN). However, it is usual to add to $\eta$, as an additional
parameter, the so-called {\it effective number of neutrinos},
$N_{\rm eff}$, which measure the amount of relativistic d.o.f. at
the time of BBN.

\section{Standard BBN theoretical predictions versus data}

The goal of a theoretical analysis of BBN is to obtain a reliable
estimate of the model parameters, once the experimental data on
primordial abundances are known. In this Section we will consider
only the case of standard BBN, where the only two free parameters
are the value of the baryon energy density parameter $\Omega_B
h^2$ (or equivalently the baryon to photon number density, $\eta$)
and possibly, a non standard value for the relativistic energy
content during BBN. The latter, after $e^\pm$ annihilation can be
parameterized in terms of the effective number of neutrinos
\be \rho_R = \left( 1 + \frac{7}{8} \left( \frac{4}{11}
\right)^{4/3} \neff \right) \rho_\gamma \pp \label{defneff}
\ee
\begin{figure}[t]
\begin{center}
\includegraphics[width=0.45\textwidth]{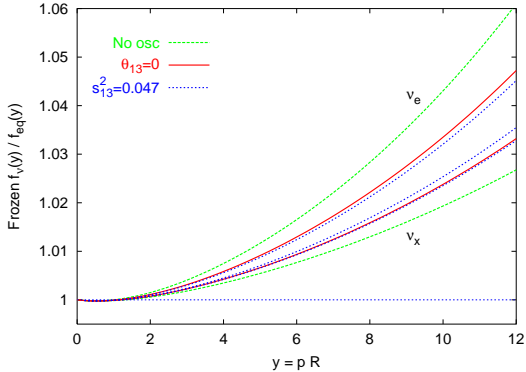}
\end{center}
\caption{Frozen distortions of the flavor neutrino spectra as a
function of the comoving momentum, for the best fit solar and
atmospheric mixing parameters. R is the scale factor. In the case
where we allow for $\theta_{13}\neq 0$ consistently with present
bounds (blue dotted lines), one can distinguish the distortions
for $\nu_\mu$ and $\nu_\tau$ (middle and lower, respectively).
From \cite{Mangano:2005cc}.} \label{fig:finalfnu}
\end{figure}

As it is well known, $\neff$ can differ from zero also {\it via}
the non thermal equilibrium terms which characterize the relic
neutrino distributions. We show in Fig.\ \ref{fig:finalfnu} the
asymptotic values of the flavor neutrino distribution, both
without oscillations and with non-zero mixing. The dependence of
the non-thermal distortions in momentum is well visible, which
reflects the fact that more energetic neutrinos were interacting
with $e^{\pm}$ for a longer period. Moreover, the effect of
neutrino oscillations is evident, reducing the difference between
the flavor neutrino distortions. Fitting formulae for these
distributions are available in \cite{Mangano:2005cc}.
\begin{table}[t]
\begin{center}
\begin{tabular}{l|l|l}
\hline\hline
Case      & $\neff$         & $\,\,\,\,\,\Delta Y_p$ \\
\hline\hline
No mixing (no QED) & 3.035 & $1.47{\times} 10^{-4}$\\
\hline
No mixing  & 3.046 & $1.71{\times} 10^{-4}$\\
\hline
Mixing, $\theta_{13}=0$ & 3.046 & $2.07{\times} 10^{-4}$\\
\hline
Mixing, $\sin^2(\theta_{13})=0.047$ & 3.046 & $2.12{\times} 10^{-4}$\\
\hline Mixing, Bimaximal, $\theta_{13}=0$ & 3.045 & $2.13{\times}
10^{-4}$ \\
\hline\hline
\end{tabular}
\end{center}
\caption{$\neff$ and $\Delta Y_p$ obtained for different cases,
with and without neutrino oscillations, as reported in
\cite{Mangano:2005cc}.} \label{tab:neffyp}
\end{table}
In Table \ref{tab:neffyp} we report the effect of non
instantaneous neutrino decoupling on the radiation energy density,
$\neff$, and on the $^4$He mass fraction. By taking also into
account neutrino oscillations, one finds a global change of
$\Delta Y_p\simeq 2.1{\times} 10^{-4}$ which agrees with the
results in \cite{Han01} due to the inclusion of QED effects.
Nevertheless the net effect due to oscillations is about a factor
3 smaller than what previously estimated, due to the failure of
the momentum-averaged approximation to reproduce the true
distortions.

In particular, the $\neff$ reported in Table \ref{tab:neffyp} is
the contribution of neutrinos to the whole radiation energy
budget, but only at the very end of neutrino decoupling. Hence,
not all the $\Delta \neff$ there reported will be really
contributing to BBN processes. In order to clarify this subtle
point, we report in Table \ref{tab:nuclides} the effect on all
light nuclides, of the non instantaneous neutrino decoupling in
the simple scenario of no neutrino oscillation, and compare this
column with the simple {\it prescription} of adding a fix $\Delta
\neff = 0.013$ contribution to radiation. Even though $Y_p$ is
reproduced (by construction), this is not the case for the other
nuclear yields.
\begin{table}[t]
\begin{center}
\begin{tabular}{l|l|l}
\hline\hline
Nuclide        & Exact         & Fixed\\
        &(No $\nu$-oscillations)          & $\Delta \neff = 0.013$\\
\hline\hline
$\Delta Y_p$   & $1.71{\times} 10^{-4}$ & $1.76{\times} 10^{-4}$\\
\hline
$\Delta$($^2$H/H) & $-0.0068{\times} 10^{-5}$ & $+0.0044{\times} 10^{-5}$\\
\hline
$\Delta$($^3$He/H) & $-0.0011{\times} 10^{-5}$ & $+0.0007{\times} 10^{-5}$\\
\hline
$\Delta$($^7$Li/H) & $+0.0214{\times} 10^{-10}$ & $-0.0058{\times} 10^{-10}$\\
\hline\hline
\end{tabular}
\end{center}
\caption{Comparison of the exact BBN results with a fixed-$\Delta
\neff$ approximation. From \cite{Mangano:2005cc}.}
\label{tab:nuclides}
\end{table}
Similar analysis have been recently presented by various groups,
which might be slightly different depending on the adopted values
of $Y_p$ and/or $^2$H/H experimental determination, see e.g.
\cite{Lis99,Bur99b,Esposito:2000hh,Bar03a,Cuo04,Cyb03c,Cyb04,Cyb05,Han02,Man07,Sim08b}.

\begin{figure}[t]
\begin{center}
\epsfig{file=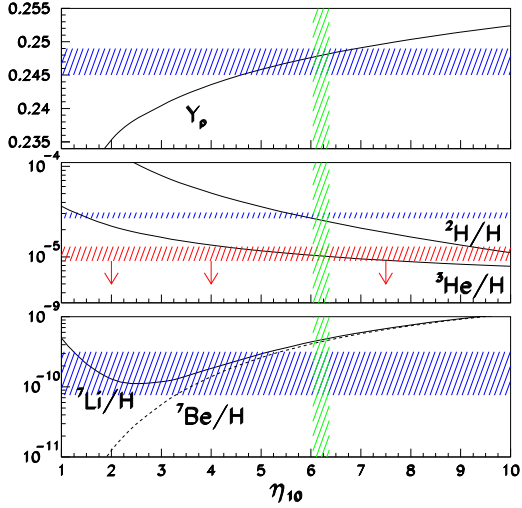,width=0.45\textwidth}
\end{center}
\caption{Values of the primordial abundances as a function of
$\eta_{10}$, calculated for $\Delta \neff = 0$. The hatched blue
bands represent the experimental determination with $1-\sigma$
statistical errors on $Y_p$, $^2H$, and $^7$Li, while the red band
is the upper bound obtained in Ref. \cite{Ban02}. Note that for
high value of $\eta_{10}$  all $^7$Li comes from $^7$Be
radioactive decay via electron capture. The vertical green band
represents WMAP 5-year result $\Omega_B h^2 = 0.02273 \pm 0.00062$
\cite{Dun08}.} \label{f:abund_parth}
\end{figure}

In the minimal BBN scenario the parameters reduces to the baryon density
only, since $\Delta \neff$ is just produced by the above non thermal
distortions. Fig. \ref{f:abund_parth} shows the dependence on
$\eta_{10}\equiv \eta \cdot 10^{10}$ of the final value of the primordial
yields, calculated using the nucleosynthesis code {\texttt{PArthENoPE}
\cite{Par08}, along with the experimental values of the abundances and
their corresponding uncertainties, as discussed in details in Ref.
\cite{Iocco:2008va}. Just to summarize, according to the analysis of
\cite{Iocco:2008va} the inferred primordial abundances result to be
\be  ^2{\rm H/H}=\left(2.87^{+0.22}_{-0.21} \right) \times 10^{-5}
\label{Dfin2} \vv \ee
\be \he3/\mbox{H} < (1.1 \pm 0.2)\times 10^{-5} \vv
\label{upperhe3} \ee
\be \yp = 0.247 \pm 0.002_{\rm stat} \pm 0.004_{\rm syst} \vv
\label{he4est} \ee
\be \left( \frac{\li7}{\rm H}\right) = \left(1.86^{+1.30}_{-1.10}
\right) \times 10^{-10}\pp \label{li7ave2} \ee
To get confidence intervals for $\eta$, one construct a likelihood
function \be {\mathcal{L}}(\eta)\propto \exp\left(
-\chi^2(\eta)/2\right) \vv \ee with \be \chi^2(\eta) = \sum_{ij} [
X_i(\eta) - X_i^{obs} ] W_{ij}(\eta) [ X_j(\eta) - X_j^{obs} ] \pp
\ee The proportionality constant can be obtained by requiring
normalization to unity, and $W_{ij}(\eta)$ denotes the inverse
covariance matrix, \be W_{ij}(\eta) = [ \sigma_{ij}^2 +
\sigma_{i,exp}^2 \delta_{ij} + \sigma_{ij,other}^2 ]^{-1} \vv \ee
where $\sigma_{ij}$ and $\sigma_{i,exp}$ represent the nuclear
rate uncertainties and experimental uncertainties of nuclide
abundance $X_i$, respectively (we use the nuclear rate
uncertainties as in Ref. \cite{Ser04b}), while by
$\sigma_{ij,other}^2$ we denote the propagated squared error
matrix due to all other input parameter uncertainties ($\tau_n$,
$G_{\rm N}$, etc.).
\begin{figure}[t]
\begin{center}
\epsfig{file=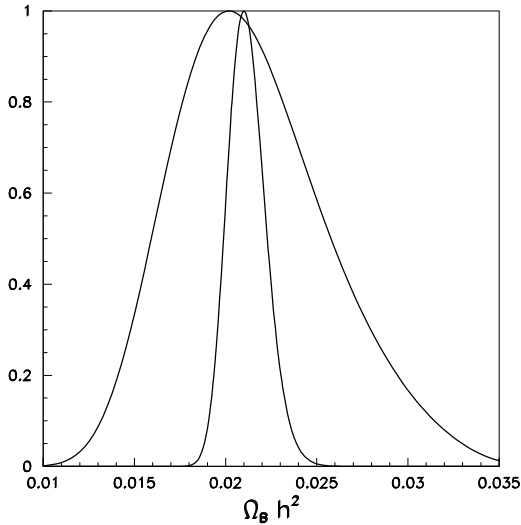,width=0.45\textwidth}
\end{center}
\caption{Likelihood functions for $^2$H/H (narrow) and $Y_p$
(broad), using only the statistical error for the $^4$He
measurement.} \label{f:like_uni}
\end{figure}
\begin{table}[b]
\begin{center}
\begin{tabular}{ccccc}
\hline\hline $\Omega_B h^2$        & 0.017  & 0.019  & 0.021  &
0.023  \\ \hline\hline $^2$H/H (10$^{-5}$)   & 4.00   & 3.36   &
2.87   & 2.48   \\ \hline $^3$He/H (10$^{-5}$)  & 1.22   & 1.14 &
1.07   & 1.01   \\ \hline $Y_p$                 & 0.2451 & 0.2462
& 0.2472 & 0.2481 \\ \hline $^6$Li/H (10$^{-14}$) & 1.72 & 1.45
& 1.25   & 1.08   \\ \hline $^7$Li/H (10$^{-10}$) & 2.53 & 3.22
& 3.99   & 4.83   \\ \hline
$^7$Be/H (10$^{-10}$) & 2.15   & 2.89   & 3.69   & 4.56   \\
\hline\hline
\end{tabular}
\end{center}
\caption{The theoretical values of the nuclear abundances for some
value of $\Omega_B h^2$.} \label{table:sbbn}
\end{table}

We first consider $^2$H abundance alone, to illustrate the role of
deuterium as an excellent baryometer. In this case the best fit values
found are $\Omega_B h^2 = 0.021\pm 0.001$ ($\eta_{10} = 5.7 \pm 0.3$) at
68\% C.L. , and $\Omega_B h^2 = 0.021 \pm 0.002$  at 95\% C.L. . A similar
analysis can be performed using $^4$He. In case only statistical error is
considered we get $\Omega_B h^2 = 0.021^{+0.005}_{-0.004}$ ($\eta_{10} =
5.7^{+1.4}_{-1.1}$) at 68\% C.L. , and $\Omega_B h^2 =
0.021^{+0.010}_{-0.006}$ at 95\% C.L. . Fig. \ref{f:like_uni} shows the
two likelihood profiles, which nicely agree. When accounting for the
largest possible systematic error on $Y_p$, the determination of $\Omega_B
h^2$ becomes even more dominated by deuterium. In any case, the result is
compatible at 2-$\sigma$ with WMAP 5-year result $\Omega_B h^2 = 0.02273
\pm 0.00062$ \cite{Dun08}. The slight disagreement might have some impact
on the determination from CMB anisotropies of the primordial scalar
perturbation spectral index $n_s$, as noticed in \cite{Pet08}, where the
BBN determination of $\Omega_B h^2$ from deuterium is used as a prior in
the analysis of the five year data of WMAP.

In Table \ref{table:sbbn} we report the values of some relevant
abundances for three different baryon densities, evaluated using
\texttt{PArthENoPE} \cite{Par08}. Notice the very low prediction
for $^6$Li and that, for these values of baryon density, almost
all $^7$Li is produced by $^7$Be via its eventual electron capture
process.

\begin{figure}[t]
\begin{center}
\epsfig{file=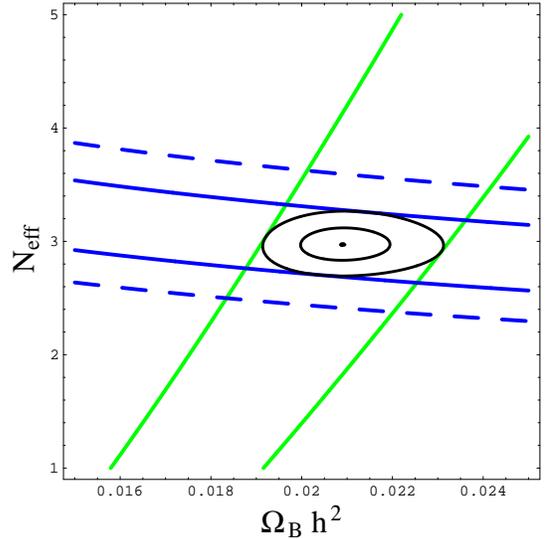,width=0.45\textwidth}
\end{center}
\caption{Contours at 68 and 95 \% C.L. of the total likelihood
function for deuterium and $^4$He in the plane ($\Omega_B
h^2$,$\neff$). The bands show the 95\% C.L. regions from deuterium
(almost vertical) and Helium-4 (horizontal), neglecting possible
systematic uncertainty on $Y_p$. We also show the 95 \% C.L.
allowed region from $Y_p$ with systematic error included (dashed
lines).} \label{f:omega_nnu_contour}
\end{figure}

If one relaxes the hypothesis of a standard number of relativistic
degrees of freedom, it is possible to obtain bounds on the largest
(or smallest) amount of radiation present at the BBN epoch, in the
form of decoupled relativistic particles, or non standard features
of active neutrinos. Fig. \ref{f:omega_nnu_contour} displays the
contour plot of the total likelihood function, in the plane
($\Omega_B h^2$,$\neff$), beautifully centered extremely close to
the standard value $\neff = 3.0$. After marginalization one gets
$\Omega_B h^2 = 0.021\pm 0.001$ and $\neff = 2.97\pm 0.14$ at 68\%
C.L., and $\Omega_B h^2 = 0.021\pm 0.002$ and $\neff =
2.97^{+0.29}_{-0.27}$ at 95\% C.L. . Note that in this case we are
using for $Y_p$ the statistical uncertainty only. More
conservatively, if one also considers the systematic uncertainty
on $Y_p$, the allowed range for $\neff$ becomes broader, $\neff =
3.0 \pm 0.3_{\rm stat}(2 \, \sigma) \pm 0.3_{\rm syst}$.

\section{Conclusions}

The ``classical parameter" constrained by BBN is the baryon to
photon ratio, $\eta$, or equivalently the baryon abundance,
$\Omega_B h^2$. At present, the constraint is dominated by the
deuterium determination, and we find $\Omega_B h^2=0.021\pm
0.001$(1 $\sigma$). This determination is consistent with the
upper limit on primordial $^3$He/H (which provides a lower limit
to $\eta$), as well as with the range selected by $^4$He
determinations, which however ever neglecting systematic errors
provides a constraint 5 times weaker. The agreement within 2
$\sigma$ with the WMAP determination, $\Omega_B h^2=0.02273\pm
0.00062$, represents a remarkable success of the Standard
Cosmological Model. On the other hand, using this value as an
input, a factor $\sim 3$ discrepancy remains with $^7$Li
determinations, which can hardly be reconciled even accounting for
a conservative error budget in both observations and nuclear
inputs. Even more puzzling are some detections of traces of $^6$Li
at a level far above the one expected from Standard BBN. Both
nuclides indicate that either their present observations do not
reflect their primordial values, and should thus be discarded for
cosmological purposes, or that the early cosmology is more
complicated and exciting than the Standard BBN lore. Neither a
non-standard number of massless degrees of freedom in the plasma
(parameterized via $\neff$) or a lepton asymmetry $\xi_e$ (all
asymmetries assumed equal) can reconcile the discrepancy. Current
bounds on both quantities come basically from the $^4$He
measurement, $\neff=3.0\pm 0.3_{\rm stat}\,(2\,\sigma)\pm 0.3_{\rm
syst}$ and $\xi_e=0.004\pm 0.017_{\rm stat}\,(2\,\sigma)\pm
0.017_{\rm syst}$.

\end{document}